# Physical interpretation of the spectral approach to delocalization in infinite disordered systems

**E G Kostadinova, C D Liaw, L S Matthews, and T W Hyde**
Baylor University

E-mail: Eva_Kostadinova@baylor.edu, Constanze_Liaw@baylor.edu, Lorin_Matthews@baylor.edu, Truell_Hyde@baylor.edu

**Abstract.** In this paper we introduce the spectral approach to delocalization in infinite disordered systems and provide a physical interpretation in context of the classical model of Edwards and Thouless. We argue that spectral analysis is an important contribution to localization problems since it avoids issues related to the use of boundary conditions. Applying the method to 2D and 3D numerical simulations with various amount of disorder $W$ shows that delocalization occurs for $W \leq 0.6$ in 2D and for $W \leq 5$ for 3D.

## 1. Introduction

In 1958 P.W. Anderson [1] suggested that sufficiently large disorder in a semi-conductor could lead to spatial localization of electrons, called Anderson localization. The subject of Anderson localization has grown into a rich field over the past decades, which led to the development of numerous definitions for localization, including the most common ones: the dynamical and statistical definitions. In the dynamical sense, an exponential decay of a wave function with respect to time implies localization of the particle represented by that wave function. In the statistical sense, localization occurs when the eigenvalues of the system's Hamiltonian (represented by a large, finite-sized, random matrix) are discrete and infinitely close to one another. The dynamical and statistical methods can easily be related to the physical concepts of scaling and perturbation theory and have been widely used as localization criteria by both physicists [2]–[5] and mathematicians [6]–[8].

A third, less popular among physicists, approach to the Anderson localization problem is found in spectral theory [9]–[12]. There, extended states are interpreted as scattering states of infinite size systems and a nontrivial absolutely continuous part of the spectral measure of a given Hamiltonian corresponds to transport. In 2013 Liaw [13] developed a spectral method, suited for detection of delocalization in infinite disordered systems of any dimension. This model has been well explained in the language of mathematics but has yet to be understood and adopted by the physics community.

The primary goal of this paper is to provide a physical interpretation of Liaw's spectral approach to delocalization problems and show its advantages and applicability to physical experiments. Here we argue that this model applied to infinite systems of any dimension will produce more accurate results than will the dynamical and statistical methods since it avoids the issues related to the use of boundary conditions. We also believe that spectral theory can improve analysis of the critical 2D Anderson localization case, where previous experimental results [14]–[17] have shown disagreement with scaling theory predictions [18].

We start with an overview of the classical approach to Anderson localization developed by Edwards and Thouless (Section 2). Next, we introduce the idea behind the spectral method and its physical interpretation (Section 3) and make comparison with the classical approach (Section 4). Then, we re-visit and analyze numerical results for 2D, and 3D systems (Section 5). Finally, we discuss future work and the spectral method's applicability in material science (Section 6).

## 2. Edwards and Thouless's Model

In 1972, Edwards and Thouless [19] developed a scaling criterion for Anderson localization and applied it to 2D and 3D numerical experiments. Their model is essentially an eigenstate-eigenvalue theory, where the sensitivity of the energy eigenvalues to the choice of periodic or antiperiodic boundary conditions is used as a criterion for localization; i.e. they argued that a particle in a localized state would not be sensitive to a change in boundary conditions. Mathematically, this can be classified as a statistical approach to localization.

### 2.1. Hamiltonian

Edwards and Thouless assumed that the wave function for an electron moving through a periodic array of sites could be described by a linear combination of site orbitals (with one orbital per site), where the amplitude for an eigenstate of energy $E$ is

$$Ea_i = -V\sum_l a_{i+l} + \epsilon_i a_i \qquad (1)$$

with the Hamiltonian at site $i$ given by

$$H_i = -ZV + \epsilon_i. \qquad (2)$$

Here $l$ is a displacement vector ranging over the $Z$ nearest neighbors of site $i$, $V$ is the coupling between sites, which is taken to be constant between neighboring sites and zero otherwise, $a_i$ is the amplitude associated with the $i^{th}$ state, and $\epsilon_i$ is the energy of the orbital localized on site $i$. With these definitions, the first term of the Hamiltonian represents the kinetic energy (or the Laplacian) and the second term the potential energy of the $i^{th}$ particle. Therefore, we can rewrite (2) in a more general operator form as

$$H_\epsilon = -\Delta + \sum_{i\in\Gamma} \epsilon_i \delta_i \langle\delta_i|, \qquad (3)$$

where $\Delta$ is the Laplacian describing a $d$-dimensional crystal with atoms located at the integer lattice points $\Gamma$ [1], and $\delta_i$ assumes the value 1 in the $i^{th}$ entry and zero in all other entries. Thus, equation (3) represents the discrete random Schrödinger operator.

In order to apply this Hamiltonian to their numerical simulations, Edwards and Thouless assumed that the $\epsilon_i$ are distributed uniformly in the interval $(-\frac{1}{2}W, \frac{1}{2}W)$, where $W$ represents the amount of disorder in the system. Thus, for any given $W$ value they could generate the $\epsilon_i$ ($1 \leq i \leq N$, where $N$ is the number of sites), determine the number of nearest neighbors $Z$ (based on the dimension and geometry of the lattice under study), and fix the constant value of the coupling $V$ (for simplicity $V$ was taken to be unity). Next, the Hamiltonian was applied to Equation (1) and the behavior of the energy eigenvalues was used as a criterion for localization.

---

[1] For the sake of simplicity we will write $\Gamma$ instead of $\Gamma^d$ to indicate a $d$-dimensional crystal.

## 2.2. Localization criterion and results

Mathematically, Edwards and Thouless's simulations aimed to solve a site percolation problem, where sites with energy $\epsilon_i < E$ are available for a particle with probability $p$ and sites with energy $\epsilon_i > E$ are unavailable for a particle with probability $1 - p$. Thus, for a given $p$, localization is dependent on the probability that a path exists for a particle to move across the crystal, where the critical value of $p$ is determined by the amount of system disorder, $W$. It is expected that as the size of the disordered system increases, it will become less likely for the particle to find a path across the crystal and eventually localization will be reached.

Edwards and Thouless's approach to the problem utilizes the sensitivity of the wave function to a change in the periodicity of the boundary conditions among sites. They calculated the energy levels for a particular system of $N$ sites first using periodic boundary conditions, and then repeated the calculation using antiperiodic boundary conditions across one of the boundaries. Next, they computed the energy difference $\Delta E$ between corresponding energy levels (when the levels are arranged in order of energy). The ratio of the energy shift $\Delta E$ to the energy spacing (roughly given by $W/N$) was used as a measure of localization. If the ratio decreases as the size of the system is increased (i.e. $\Delta E$ is small when compared to the energy spacing) the states are localized, but if it does not decrease ($\Delta E$ is of the order of the energy spacing) the states are delocalized.

Edwards and Thouless showed that Anderson's localization criterion can be relaxed for both the 2D and 3D cases. Their simulations, along with later work by Thouless [20], indicated that a transition from extended to localized states will occur for $W \geq 6$ in the 2D case and for $W > 15$ in the 3D case. However, the 3D simulation suggested that for $5 \leq W \leq 15$ the behavior of the wave functions was characteristic neither of exponential decay, nor of extended states. It is reasonable to conclude that the transition point distinguishing localized from extended states is located in the interval $W \in [5,15]$. Further study by Last and Thouless [21] gave support to the notion that the solutions in this interval might be localized but falling off with a power law instead of exponentially. Thouless argued that similar phenomenon occurs in 2D but the effect is less prominent and harder to observe [20].

## 2.3. Discussion

Edwards and Thouless's work laid the grounds for the development of scaling theory by Abrahams [18], which explains how transport and size of the medium are related. Scaling has been widely used in analysis of localization problems and has given repeatable results in agreement with theoretical predictions in both 1D [22] and 3D systems [2], [23], [24]. Nevertheless, there is still disagreement between theory and experiment for the 2D case, where Anderson localization is considered to be a critical phenomenon. According to Abrahams, in 2D systems conductance always decreases with size increase and localization can always be reached. However, extended states in 2D have been observed in electron structures [14], MOSFETs [15], and other 2D systems [16], [25]. In addition, experiments with light [26], [27] and atomic kicked rotors [28] have shown that localization is a function of the disorder strength and that transport can occur at weak disorder [26].

We argue that some of the disagreement observed in the 2D case may arise from the use of periodic boundary conditions in scaling theory calculations and the notion that an increase in size can be accurately achieved by taking a system of $N$ sites as a unit cell and then periodically extending it to obtain an infinitely periodic system. These and other issues can be avoided using the suggested spectral approach.

## 3. The spectral approach

This paper does not aim to contradict scaling theory. Rather, its goal is to introduce an alternative mathematical approach to Anderson localization problems that provides additional insight into the theory while improving agreement between numerical and experimental results. In this section, we start with a brief overview of cyclicity and the Spectral Theorem, before introducing Liaw's spectral method in mathematical terms. We then develop a physical interpretation of the model that will allow comparison with Edwards and Thouless's method.

### 3.1. Definition of Cyclicity

An operator $A$ on a Hilbert space $\mathcal{H}$ has a cyclic vector $f$ if the span of the vectors $\{f, Af, A^2f, ...\}$ is dense in $\mathcal{H}$. Equivalently, $f$ is a cyclic vector for $A$ when the set of all vectors of the form $p(A)f$, where $p$ varies over all polynomials, is dense in $\mathcal{H}$. In order to examine this idea, we will assume $\mathcal{H}$ to be finite dimensional in this section.

If $A$ is a finite Hermitian matrix, then a necessary and sufficient condition that $A$ has a cyclic vector is that its eigenvalues are distinct. If they are not distinct, nothing is sufficient to make any vector $f$ cyclic.

In terms of quantum mechanics, let us consider the (time-independent) Hamiltonian $H$ on a Hilbert space $\mathcal{H}$. This Hamiltonian represents a finite-dimensional discrete operator in $\mathcal{H}$ with finitely many eigenvalues $\lambda_i$. It is immediately clear that if $H$ has (mutually) nondegenerate eigenvalues, $\lambda_i \neq \lambda_j$ when $i \neq j$, then $H$ is cyclic in the Hilbert space. On the contrary, if $H$ has degenerate eigenvalues, then it is not a cyclic operator in the Hilbert space.

To see this mathematically, we begin with a few observations. Consider the eigenvector-eigenvalue equation:

$$Hv = \lambda v. \qquad (4)$$

We see that the sequence of vectors $\{v, Hv, H^2v, ...\}$ consists of vectors that are all parallel since $H^2v = \lambda^2 v, H^3v = \lambda^3 v, ...$, i.e. each vector in the sequence is a constant multiple of $v$.

If two vectors are parallel, then any linear combination of them must be parallel to both. The span of two or more parallel vectors consists of the line through the origin which contains these vectors. Apparently, such a series cannot constitute a basis in Hilbert space.

Therefore, we can conclude that $v$ is not a cyclic vector in $\mathcal{H}$. In particular:

$$span\{v, Hv, H^2v, ...\} = span\{v\} \neq \mathcal{H}. \qquad (5)$$

Suppose $H$ has degenerate eigenvalues for the eigenvectors $v_1$ and $v_2$, that is, $\lambda_1 = \lambda_2 = \lambda$. Then for any linear combination of $v_1$ and $v_2$, $v_3 = av_1 + bv_2$, we have

$$Hv_3 = H(av_1 + bv_2) = \lambda_1 av_1 + \lambda_2 bv_2 = \lambda(av_1 + bv_2) = \lambda v_3. \qquad (6)$$

Thus, a vector equal to a linear combination of two eigenstates having degenerate eigenvalues will generate a sequence of parallel vectors that cannot span the Hilbert space.

This is also true for any general vector $v$ in $\mathcal{H}$ (where $v$ is not necessarily a linear combination of eigenvectors of $H$). Mathematically, the span of the eigenvectors, $span\{v_1, v_2\}$, constitutes a subspace embedded into the Hilbert space. Therefore, the parts of $\{v, Hv, H^2v, ...\}$ that lie within $span\{v_1, v_2\}$ are parallel and cannot span the two dimensional subspace, so the generated sequence will not span the Hilbert space. This means that when the Hamiltonian has degenerate eigenvalues the general vector $v$ and the operator $H$ are not cyclic in $\mathcal{H}$.

On the other hand, if all eigenvalues are nondegenerate, $\lambda_1 \neq \lambda_2 \neq \lambda_3 ...$ then any vector in $\mathcal{H}$ can be expressed as: $\tilde{v} = \sum_{i=1}^{n} c_i v_i$ for $c_i \neq 0$, where the sequence $\{\tilde{v}, H\tilde{v}, H^2\tilde{v}, ...\}$ now consists of vectors that are linearly independent. This means that the vector $\tilde{v}$ is cyclic for the operator $H$ in $\mathcal{H}$. In particular:

$$span\{\tilde{v}, H\tilde{v}, H^2\tilde{v}, ...\} = \mathcal{H}. \tag{7}$$

Let us now discuss how to extend from this finite dimensional scenario to operators in infinite dimensional Hilbert spaces.

*3.2. The Spectral Theorem*

In the case of a quantum mechanical lattice $\Gamma$, the Hilbert space of interest is the square-summable space $l^2(\Gamma)$ which consists of the square-summable vectors with entries on the lattice points (i.e., the sum of the entries squared is finite).

Recall that, if the Hamiltonian has only a finite number of distinct eigenvalues, the possible vector states for the particle are either the corresponding eigenvectors or some combination of them. Therefore, possible solutions for the wave function will all belong to the square-summable Hilbert space of the lattice $l^2(\Gamma)$. The Hamiltonian can then be diagonalized by a change of basis

$$H = U^{-1}DU, \text{ where } D = \begin{pmatrix} \lambda_1 & \cdots & 0 \\ \vdots & \ddots & \vdots \\ 0 & \cdots & \lambda_n \end{pmatrix}, \tag{8}$$

where $U$ is the unitary matrix that contains the orthonormalized eigenvectors in its columns.

The generalization of this procedure in operator theory is called the Spectral Theorem. Broadly speaking, the Spectral Theorem provides the conditions under which a general operator or a matrix can be diagonalized. It identifies a class of linear operators that can be modeled by multiplication operators.[2]

For the case in hand, consider a more general Hamiltonian $H$ acting on the vectors of $l^2(\Gamma)$. If this Hamiltonian is "linear" and if $v \in l^2(\Gamma)$, then we also have $Hv \in l^2(\Gamma)$. The goal is to diagonalize $H$. The Spectral Theorem says that when $H$ is self-adjoint and cyclic, a unitary operator $U$ exists so that

$$H = U^{-1}M_\xi U, \tag{9}$$

where $M_\xi f(\xi) = \xi f(\xi)$ is the multiplication by the independent variable on another square-integrable Hilbert space $L^2(\mu)$. The new space $L^2(\mu)$ stands for the square-integrable functions with respect to $\mu$

$$L^2(\mu) = \left\{ f: \mathbb{R} \to \mathbb{R} \mid \int_\mathbb{R} |f(\xi)|^2 d\mu(\xi) < \infty \right\}. \tag{10}$$

---

[2] *In operator theory, a multiplication operator is an operator $T$ defined on some vector space of functions and whose value at a function $f$ is given by a fixed function $h(\xi)$. That is, when $T$ acts on $f$, the result is $h(\xi)f(\xi)$:*
$$T(f)(\xi) = h(\xi)f(\xi)$$
*for all $f$ in the function space and all $\xi$ in the domain of $f$ (which is the same as the domain of $h$). In equation (9) we will encounter a multiplication operator for which $h(\xi) = \xi$.*

In other words, the new square-integrable Hilbert space $L^2(\mu)$ is the set of all real-valued functions ($f: \mathbb{R} \to \mathbb{R}$) whose square integral is bounded. This definition guarantees the conservation of probability in the new space, which is needed for a proper quantum mechanical treatment of the problem. In essence, the spectral measure $\mu$ encodes all the spectral information of $H$ and can now allow for discrete and continuous values, as well as anything "in between". As a result, we can decompose $d\mu$ into two parts:

(i)     $d\mu_{ac}$ – the absolutely continuous part of the spectrum of $H$, which corresponds to the scattering states of the system or the conducting band of a semiconductor. By the RAGE Theorem [12], the existence of $d\mu_{ac} \neq 0$ means that there is delocalization in terms of transport.

(ii)     $d\mu_{sing}$ – the singular part of the spectrum of $H$ that represents "everything else", including the discrete eigenvalues, where the eigenvalues are included as Dirac $\delta$ point masses. For example, if $H$ only has one eigenvalue at $\lambda$, then $\mu$ equals a Dirac $\delta$ mass at $\lambda$:

$$\int_{\mathbb{R}} f d\mu = \int_{\mathbb{R}} f \delta_\lambda d\xi = f(\lambda). \tag{11}$$

It is important to note that this part of the spectrum of $H$ also contains very poorly behaved pieces, called the singular continuous part.[3]

We see that when the Hamiltonian not only has discrete values, but also an absolutely continuous part of its spectrum, we cannot simply use a change of basis to diagonalize it as in (8). In this case, we need the Spectral Theorem to switch from the square summable Hilbert space to the square-integrable Hilbert space whose measure accounts for all possible solutions. (Naturally, continuous solutions require an integrable space and cannot be contained in a summable one.) In order to diagonalize $H$, we need to transform $H$ on $l^2(\Gamma)$ to $M_\xi$ on $L^2(\mu)$.

*3.3. Essence of the spectral method*

Given a Hamiltonian $H$ on a Hilbert space $\mathcal{H}$, consider its spectral measure $\mu$. We can now decompose this measure into an absolutely continuous part and a singular part:

$$\mu = \mu_{ac} + \mu_{sing}, \tag{12}$$

where the space $L^2(\mu)$ (on which $H$ acts by $M_\xi$) can be decomposed into two orthogonal Hilbert spaces:

$$L^2(\mu) = L^2(\mu_{ac}) \oplus L^2(\mu_{sing}). \tag{13}$$

where the symbol $\oplus$ is the direct product of the two subspaces $L^2(\mu_{ac})$ and $L^2(\mu_{sing})$. In other words, every element $f$ from the space $L^2(\mu)$ can be written as $f = f_{ac} + f_{sing}$, where $f_{ac}$ is an element from $L^2(\mu_{ac})$ and $f_{sing}$ is an element from $L^2(\mu_{sing})$. Conversely, every element $g$ from $L^2(\mu_{ac})$ and every element $h$ from $L^2(\mu_{sing})$ are orthogonal ($\int g(\xi)h(\xi)d\mu(\xi) = 0$). In this case the Hamiltonian has a part $H_{ac}$ that comes from $L^2(\mu_{ac})$ and a part $H_{sing}$ that comes from $L^2(\mu_{sing})$.

---

[3] *It is not known what physical property/state corresponds to the singular continuous part (see p.23 of [29]). We saw a hint for a possible answer in Edwards and Thouless' model, where they pointed out that there are values of their solution corresponding to an intermediate region in which the particle states are neither localized, nor describable in terms of weakly coupled plane waves [19]–[21]. However, these "intermediate" states found in theory have not yet been observed experimentally.*

Examining the Hamiltonian from operator theory (3) that we generalized from the Hamiltonian used in Edwards and Thouless' method we have:

$$H_\epsilon = -\Delta + \sum_{i \in \Gamma} \epsilon_i \delta_i \langle \delta_i |, \qquad (14)$$

where again the $\epsilon_i$ are distributed uniformly in the interval $(-\frac{1}{2}W, \frac{1}{2}W)$. For any vector $\delta_i \in l^2(\Gamma)$, say $\delta_0$, we can state the following theorem [11]:

<u>*Theorem*</u>: *$\delta_0$ is cyclic for the singular part $(H_\epsilon)_{sing}$ almost surely.*

The main idea of the spectral method is that if we can show that $\delta_0$ is *not* cyclic for $H_\epsilon$, then almost surely

$$(H_\epsilon)_{sing} \neq H_\epsilon. \qquad (15)$$

If (15) holds, we can conclude that there is more to $H_\epsilon$ than the singular part, which means that there is an absolutely continuous part implying the existence of extended states.

### 4. Comparison between Edwards & Thouless and Liaw

Both the statistical method of Edwards and Thouless (Section 2) and the spectral method developed by Liaw (Section 3) approach Anderson localization using the form of the Hamiltonian given by Equation (3). Furthermore, both models assume that the energies of the sites in the crystal are distributed uniformly in the interval $(-\frac{1}{2}W, \frac{1}{2}W)$. What appears to be different is the delocalization criterion and the method of mathematical calculation. However, the logic behind the two approaches is similar.

As noted, Edwards and Thouless examine the ratio of the energy shift $\Delta E$ to the energy spacing $W/N$ and argue that localization occurs if the ratio goes to zero as the size of the system increases. This is a very logical definition of localization since the ratio in question can be viewed as the ratio between the energy band width and the band gap width in a material. Naturally, if the band gap is much larger than the energy band width (i.e., $\Delta E/(W/N) \to 0$), the electrons in the material cannot easily achieve transport. On the contrary, as the band gap tends to zero, separate energy bands merge and the electrons can freely transport through the crystal. This is illustrated in figure 1.

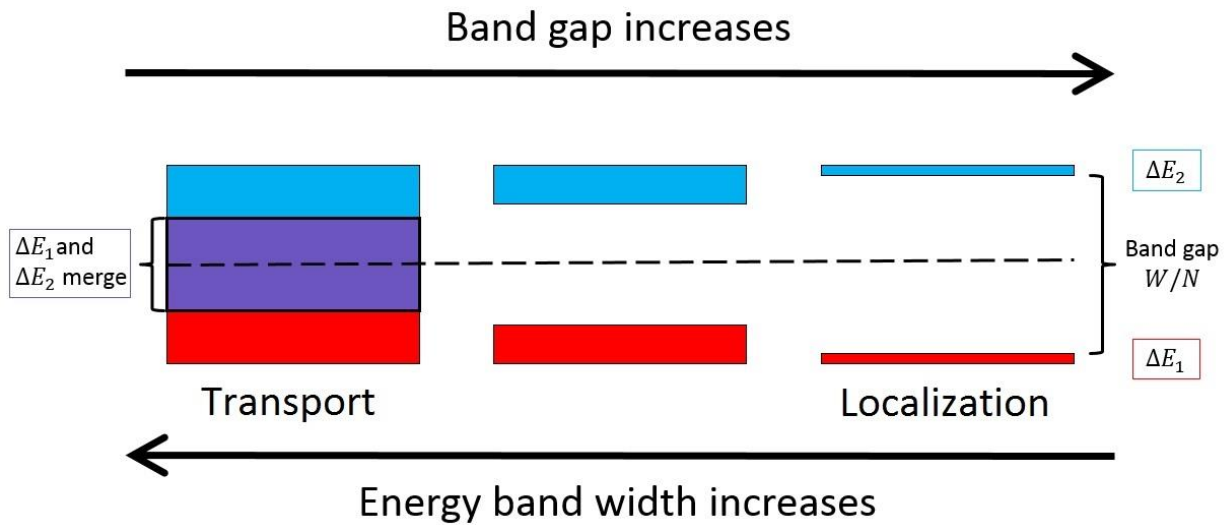

Figure 1. As the band gap decreases, the energy bands eventually merge.

In comparison, Liaw shows that if the Hamiltonian of the system exhibits a continuous (or partially continuous) spectrum, then there is delocalization. In essence, this definition of transport parallels the merger of energy bands in the Edwards and Thouless model.

Edwards and Thouless' model predicts localization whenever the band gap is much larger than the energy band width. Therefore, in this approximation, each energy band may be considered to be a discrete energy eigenvalue[4]. Mathematically, this corresponds to a pure point spectrum of the Hamiltonian (i.e., only discrete eigenvalues). In the case of a pure point spectrum, the random nature of the system ensures non-degeneracy of the eigenvalues with probability one. For example, even if two eigenvalues are equal for one realization of the random variables, the effect of slightly changing just one of these random variables will split them into distinct eigenvalues. In such a system, by the Kolmogorov zero-one law, all eigenvalues are distinct. We know from quantum mechanics that in the absence of degeneracy in the eigenvalues of the Hamiltonian, if a measured value of the energy of a quantum system is determined, the corresponding state of the system is assumed to be known, since only one eigenstate corresponds to each energy eigenvalue. Clearly, this case corresponds to localization.

On the other hand, Liaw's delocalization corresponds to a continuous part of the spectrum. The existence of an absolutely continuous spectrum shows a strong form of delocalization, the so-called dynamical delocalization [12]. In the presence of an absolutely continuous spectrum, the energy bands already overlap by a non-trivial amount, which indicates some degeneracy. In this case the effect of slightly changing the random variables induces a "wiggling" of the band edges, but does not change the overlapping property. In this sense, degeneracy (i.e. overlap) of the absolutely continuous spectrum is somewhat more stable than is the pure point spectrum case. In quantum mechanics an increasing degeneracy of the energy states indicates a decreasing probability of finding the particle at a specific state. This allows us to intuitively understand the relation between delocalization and degeneracy of the Hamiltonian under question.

Therefore, Liaw starts with the classical Anderson localization problem formulation and implements well established ideas into a rigorous new mathematical approach. The above comparison shows that the spectral approach has a logical physical interpretation and is suitable for analysis of infinite-dimensional physical systems.

So far we have emphasized that the Hamiltonian operator $H_\epsilon$ in both methods has essentially the same form (Anderson-type Hamiltonian). However, Edwards and Thouless use an *unbounded Hamiltonian*, which requires the application of boundary conditions to the wavefunction, whereas the spectral approach starts with a *bounded* $H_\epsilon$, which can be applied to the Schrödinger equation without the assumption of boundary conditions. Thus, an important distinction between the two treatments is in the domain of the applied operator.

The standard formulation of quantum mechanics and quantum field theory uses unbounded operators, which are not defined on the entire Hilbert space $\mathcal{H}$. Such operators are restricted only to dense subspaces $\mathcal{H}$. Although this approach is successful in solving simple problems in quantum mechanics, it causes incompleteness of the theory. For instance, although the dynamics of quantum systems requires the use of strictly self-adjoint operators, the densely defined unbounded operators often used in introductory texts are Hermitian but not necessarily self-adjoint. In addition, important physical properties of the self-adjoint operators are sensitive to the choice of the domain. Thus, restriction of the domain to only dense subspaces of the Hilbert space, can cause loss of information about the physical properties of the system [30].

The algebras of bounded operators has been shown to form an equivalent treatment of quantum mechanics [31] and quantum field theory [32]. Here we avoid the (rather unphysical) application of boundary conditions using an Anderson-type bounded Hamiltonian operator, which is defined on the entire Hilbert space. Thus, the proposed spectral approach can account for wavefunction solutions that are omitted by scaling theory due to the restricted domain of the unbounded Hamiltonian used there.

---

[4] *It is known that even if the energy bands are small, they can still have internal structure. However, Edwards and Thouless' model cannot distinguish structure inside the bands and for the purpose of their simulation each band behaved as a discrete energy value.*

## 5. Numerical Work

In Section 3, we showed that the existence of $i$ such that a vector $\delta_i \in l^2(\Gamma)$ is noncyclic for a given $H_\epsilon$ implies transport[5]. To accomplish this numerically, we start with the general form of the perturbed Hamiltonian (Equation (3)), where we fix the dimension $d$ and the Laplacian $\Delta = -ZV$, where $Z$ is determined by the geometry of the lattice and $V$ is taken to be unity. Then for a given value of the disorder $W$ we generate one realization of the random variables $\epsilon_i$, which are independent and identically distributed in the interval $[-W/2, W/2]$. Next we fix a random vector, $\delta_0$, in the $d$-dimensional space and generate the sequence $\{\delta_0, H\delta_0, H^2\delta_0, \ldots H^n\delta_0\}$, where $n \in \{0,1,2,\ldots\}$ corresponds to the number of iterations of the Hamiltonian and is used as a timestep. The Gram-Schmidt orthogonalization process (without normalization) is then applied to the members of the sequence and the resulting subspace is denoted by $\{m_0, m_1, m_2, \ldots, m_n\}$. If we let $\|\cdot\|_2$ denote the Euclidean norm, then the distance from $\delta_0$ to the $n$ dimensional orthogonal subspace is given by

$$D^n_{\epsilon,W} = \sqrt{1 - \sum_{k=0}^{n} \frac{\langle m_k, \delta_0 \rangle^2}{\|m_k\|_2^2}}. \tag{16}$$

It can be shown [13] that if

$$\lim_{n \to \infty} D^n_{\epsilon,W} > 0, \tag{17}$$

then $\delta_0$ is noncyclic for $H_\epsilon$ and we can conclude the existence of transport. It is important to mention that $D \leq 0$ does not necessarily imply localization; however, it is reasonable to expect that it indicates the lack of extended states. (Here $D$ has no physical meaning. Calculating its value is a mathematical technique used to test for cyclicity[6].)

Once the values of $D$ at each timestep are generated for one realization of the random variables $\epsilon_i$ (for a given $W$), we repeat the same process several times. Then the discrete time evolution of the averaged $D$ values is be plotted. (In sections 5.1 and 5.2 we will report results using 15 realizations for each $0.1 \leq W \leq 5$ in 3D, 4 realizations for each $W > 5$ in 3D, and 4 realizations for each $W$ in 2D.) In our graphs the number of iterations of the Hamiltonian is represented by a discrete timestep $n \in \{0,1,2,\ldots\}$ (i.e., one iteration corresponds to the interaction of the original state with the nearest neighbors, the second iteration corresponds to the interaction of the new state with the nearest neighbors, etc.). In this sense, to prove delocalization, it suffices to find $W > 0$ for which

$$\lim_{n \to \infty} D^n_{\epsilon,W} \neq 0. \tag{18}$$

---

[5] In accordance with [13], in order to determine if delocalization occurs for a given realization of disorder in a system, it is sufficient to choose one vector and then test for cyclicity.

[6] A derivation of equation (16) and a proof of the relationship between the D value and cyclicity in operator theory can be found in [13], [30].

However, in a physical experiment, it will be sufficient to show that as time goes on, the distance does not decrease to zero rapidly enough to affect the results.[7]

It is important to emphasize that the spectral method can be applied to infinite disordered systems of any dimension. Nevertheless, similar to Edwards and Thouless's work, here we report the results from applying the method to a 2D square lattice and a 3D diamond lattice. We used the raw data from the 2D and 3D numerical simulations conducted by Liaw et al. [13], [30] to obtain discrete time evolution plots of the distance $D$ for various values of the disorder $W$ (sections 5.1 and 5.2). We also employed equation fitting to extract critical exponents for each plot, which were then used to set a criterion for delocalization (section 5.3). In the following sections we assume weak disorder for $W < 1$, medium disorder for $1 \leq W \leq 5$, and strong disorder for $W > 5$. (Similar definitions can be found in [31], [32].)

## 5.1. Results in 2D

Here we examine 2D simulations with small values of the disorder. In figure 2 we show time evolution distance plots (figure 2a) along with corresponding log-log plots (figure 2b) for values of disorder $W = 0.1, 0.2, \ldots, 1.2$ and timestep $n = 4500$. As can be seen from the graphs, the distance parameter does not rapidly tend to zero for $W \leq 0.6$, implying the existence of extended states. For $0.6 < W < 0.9$, the slope of the lines increases slightly and it is less obvious whether $D$ will cross the zero axis at infinity. It is interesting to note that this interval may correspond to the transition area discussed by Touless [20]. As the disorder increases to $W \geq 0.9$, the slopes of the lines become even greater, suggesting that transition to localization has probably occurred. These trends can also be seen in the log-log plots (figure 2b), where the large slope of the lines for $W \geq 0.6$ implies faster decay of the distance vector towards zero.

In figure 2 we can also observe a gap between the limiting values of $D$ as the disorder increases from $W = 0.3$ to $W = 0.4$, from $W = 0.6$ to $W = 0.7$, and from $W = 0.8$ to $W = 0.9$, which may imply some type of splitting in the allowed energy states of the system. This phenomenon will be examined in detail in a future paper.

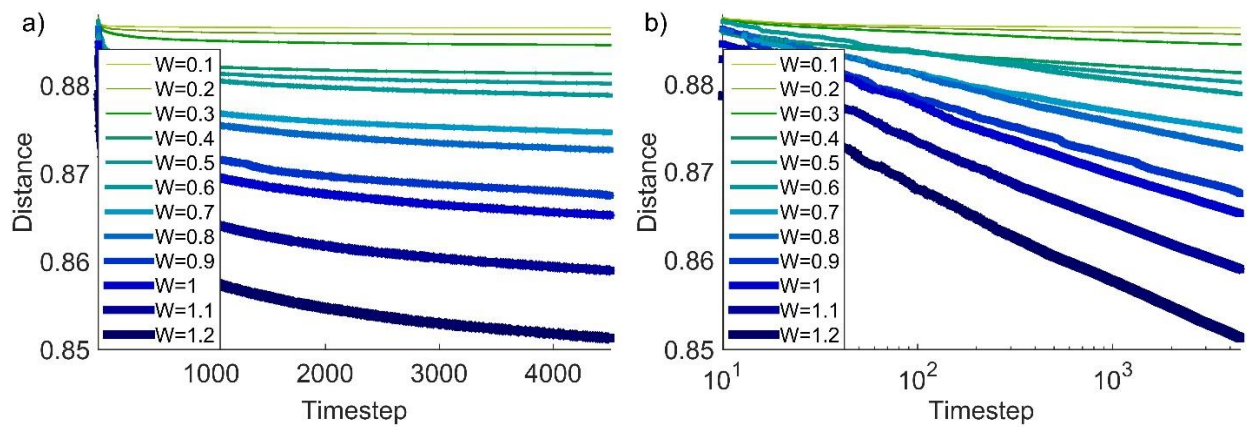

Figure 2. a) Time evolution plot of $D$ for the 2D weak disorder case. B) Corresponding log-log plot.

## 5.2. Results in 3D

In order to test the validity of the spectral model for the less controversial 3D case, we produced 3D simulations over a wide variety of $W$ values to see if our results are in agreement with Edwards and Thouless's model. This time, however, we used $n = 500$ due to data requirement restrictions. Figure 3 shows distance time evolution plots and log-log plots for three sets of disorder: weak, medium, and strong, where the perturbation ranges from $W = 0.1$ to $W = 40$. For small and medium disorder (figures 3a through 3d) the distance clearly tends to nonzero values and the corresponding log-log plots have constant slopes, implying no significant decrease of $D$ over time. In the case of large disorder, figures 3e and 3f show that for $W \geq 10$ the distance lines approach the horizontal axis more rapidly and the slopes of the corresponding log-log plots become increasingly negative. In this case the model cannot give conclusive results, which suggests that a transition to localization has occurred. The graphs also indicate that the transition point distinguishing extended from localized states occurs in the interval $5 < W < 15$. The existence of such a transition point and the interval of disorder values where it should occur are in agreement with Edward and Thouless's numerical simulations (see section 2.2.).

---
[7] *Similarly, physicists employing the classical dynamical approach define localization as an exponential decay, i.e. a decay that occurs rapidly enough to be observable in an experiment* [18].

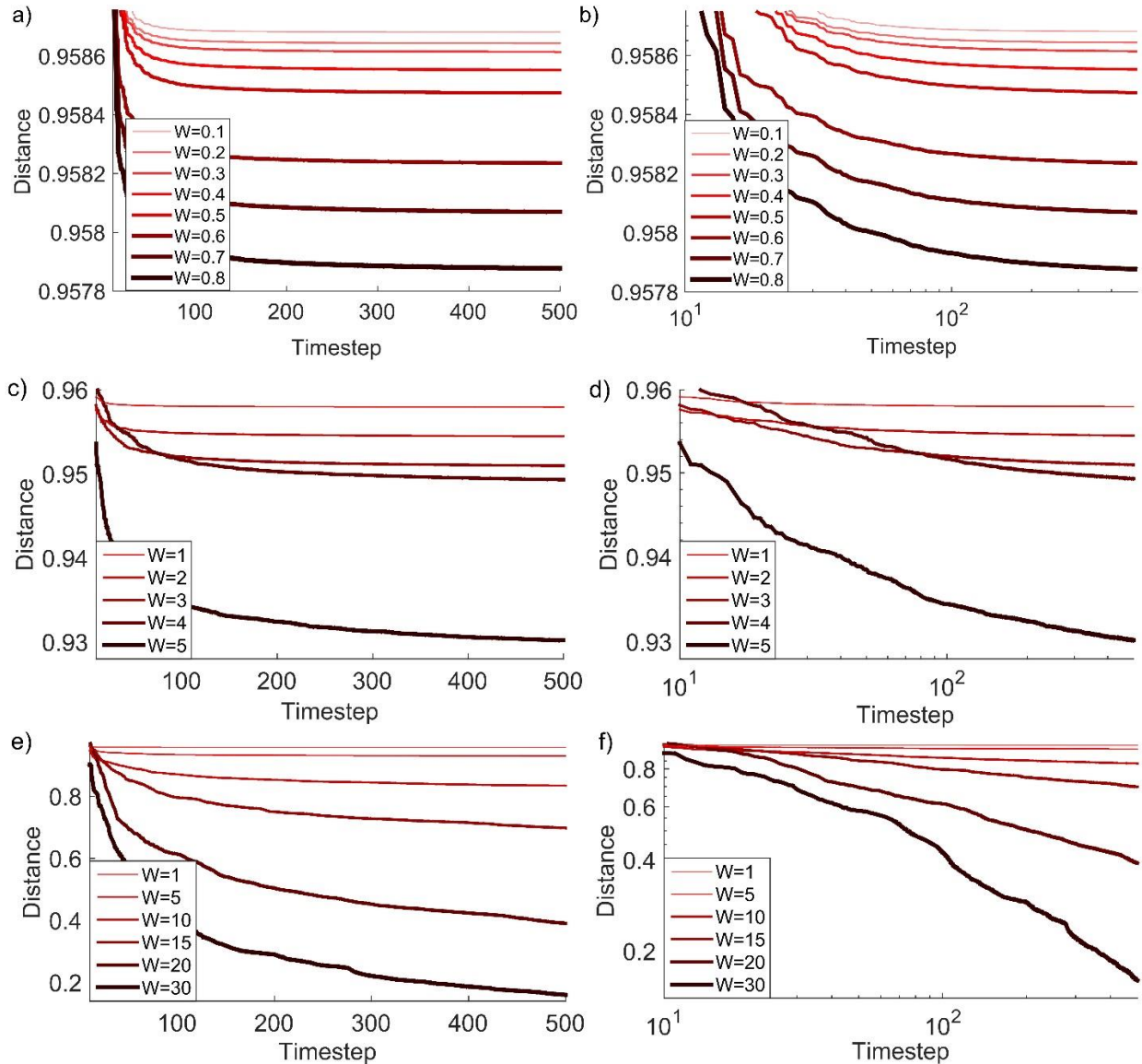

Figure 3. Time evolution plot of $D$ for a 3D case with: a) weak disorder, c) medium disorder, and e) strong disorder. Graphs b), d), and f) show the corresponding log-log plots.

*5.3. Equation fitting*

To determine if the obtained $D$ values approach a finite positive number at infinity, implying the existence of extended states, we calculated the parameters necessary to fit the data using the equation

$$y = mx^{-\alpha} + b, \qquad (19)$$

where $\alpha$ (the critical exponent) indicates how fast the distance term tends to a finite value and $b$ (the $y$-intercept) corresponds to the limiting value of $D$ as $n \to \infty$.

To test the goodness of the fit and provide a better visual representation, we rescaled the $x$-axis by $X = x^{-\alpha}$ and performed a polynomial fitting to the linear equation $y = mX + b$. We omitted the first 200 points for 2D and the first 120 points for 3D in order to avoid skewed results due to the rapid change of $D$ in these initial intervals. Then for both 2D and 3D we generated fits for $\alpha = 0.01\!:\!0.01\!:\!2$ and analyzed the error. The value of $\alpha$ that gave the smallest norm of the residuals was then used to select the best fit for each disorder $W$ and extract the corresponding $m$ and b values (see tables 1 through 5).

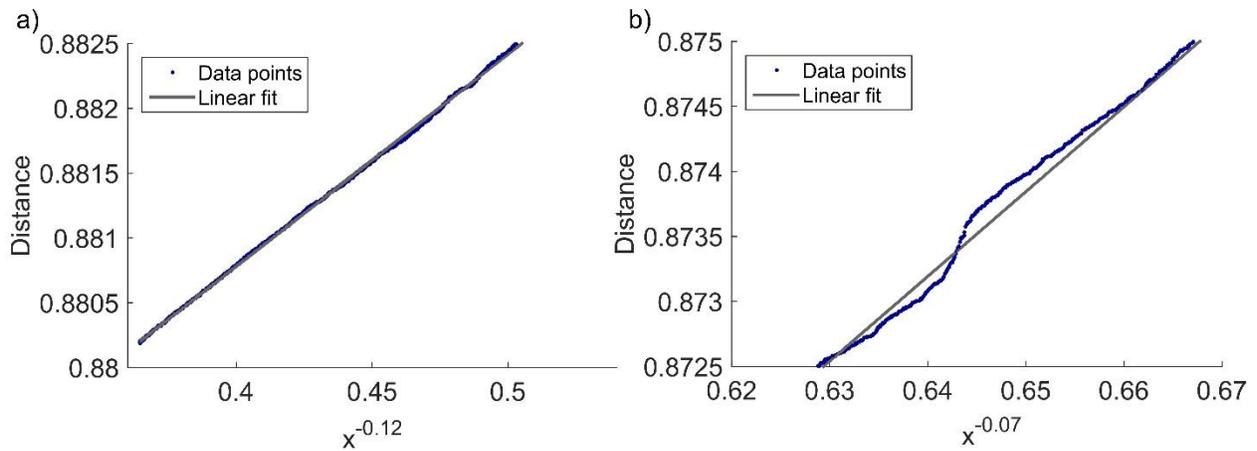

Figure 4. Best linear fit of a) 2D, $W = 0.5$, $\alpha = 0.12$ and b) 2D, $W = 0.9$, $\alpha = 0.07$. The linear fit becomes less accurate as the value of the disorder increases.

Based on linear regression analysis, we defined two further requirements necessary for establishing the existence of extended states:

(i)     $\alpha \geq 0.1$ and
(ii)    $err \leq 0.001$.

If $\alpha < 0.1$, the rescaled graphs exhibit an increasing number of concave regions and the line fit may result in overestimating the limiting distance value. The appearance of these concave regions indicates that equation (19) is not an accurate representation of the data and that the method cannot establish a nonzero $D$ at infinity.

Another criterion we used to establish the existence of extended states is the amount of error in the linear fits. If the norm of the residuals is bigger than $0.001$, we do not expect that the rescaled plot under study can be approximated by a line. Therefore, for $err > 0.001$ we believe that the original fit is not best described by equation (19) and the distance does not necessarily approach a finite nonnegative value at infinity. Figures 4a and 5a show examples of "good" linear fits for 2D and 3D respectively, whereas figures 4b and 5b show examples of linear fits that fail at least one of the requirements.

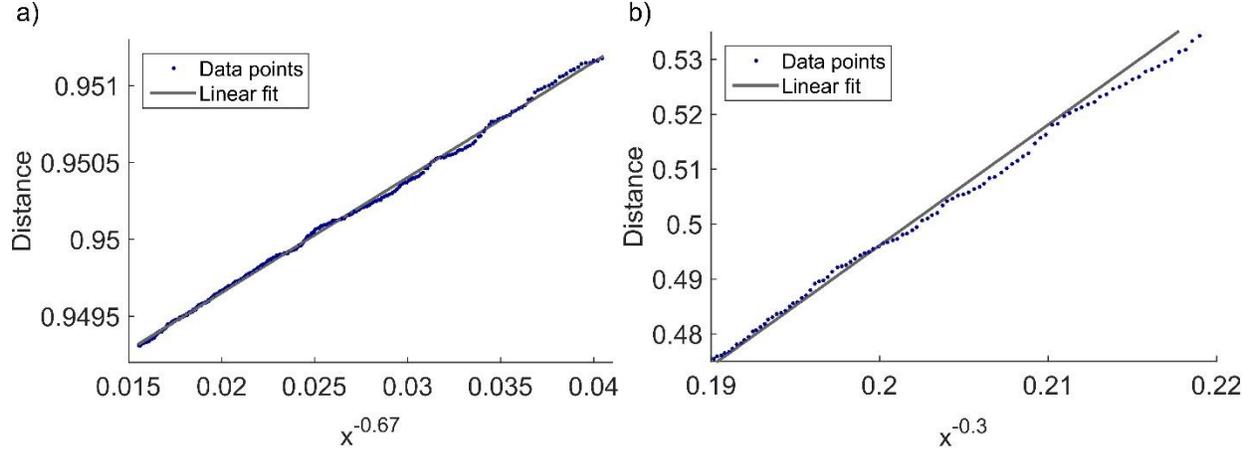

Figure 5. Best linear fit of a) 3D, $W = 4$, $\alpha = 0.67$ and b) 3D, $W = 20$, $\alpha = 0.3$. The linear fit becomes less accurate as the value of the disorder increases. Notice that the $y$-scale in a) is finer than in b).

Applying this analysis to the data in table 1 through 5 indicates that delocalization occurs in 2D for $W \leq 0.6$ and in 3D for $W \leq 5$. In the 2D case, tables 1 and 2 show that for $W \geq 0.7$, the critical exponent is smaller than the accepted minimum and for $W \geq 0.8$ both the minimum and the error criteria are violated. In the 3D case with strong disorder, Table 5 shows that delocalization does not occur for $W \geq 10$. Although some of the calculated exponents in this case are above the required minimum, the error in the calculation is significant. The values of $\alpha$ and $err$ that violate the delocalization criteria are bolded in all tables.

Table 1. Values of $\alpha, m$, and $D$ with the corresponding error for 2D weak disorder.

| $W$ | 0.1 | 0.2 | 0.3 | 0.4 | 0.5 | 0.6 | 0.7 | 0.8 | 0.9 |
|---|---|---|---|---|---|---|---|---|---|
| $\alpha$ | 0.15 | 0.16 | 0.18 | 0.12 | 0.12 | 0.12 | **0.09** | **0.07** | **0.07** |
| m | 0.0011 | 0.0036 | 0.0070 | 0.0111 | 0.0163 | 0.0234 | 0.0327 | 0.0479 | 0.0651 |
| b | 0.8862 | 0.8848 | 0.8830 | 0.8773 | 0.8743 | 0.8704 | 0.8593 | 0.8461 | 0.8315 |
| err | 0.0001 | 0.0002 | 0.0002 | 0.0004 | 0.0006 | 0.001 | 0.001 | **0.002** | **0.005** |

Table 2. Values of $\alpha, m$, and $D$ with the corresponding error for 2D medium disorder.

| $W$ | 1 | 1.1 | 1.2 | 1.3 |
|---|---|---|---|---|
| $\alpha$ | **0.05** | **0.05** | **0.01** | **0.01** |
| $m$ | 0.0875 | 0.1031 | 0.4536 | 0.5417 |
| $b$ | 0.8078 | 0.7913 | 0.4343 | 0.3559 |
| $err$ | **0.002** | **0.002** | **0.003** | **0.004** |

Table 3. Values of $\alpha, m$, and $D$ with the corresponding error for 3D weak disorder.

| $W$ | 0.1 | 0.2 | 0.3 | 0.4 | 0.5 | 0.6 | 0.7 | 0.8 | 0.9 |
|---|---|---|---|---|---|---|---|---|---|

|   α   | 1.75    | 1.35    | 1.21    | 1.10    | 1.02    | 1.00    | 1.05    | 1.08    | 1.41    |
|-------|---------|---------|---------|---------|---------|---------|---------|---------|---------|
|   m   | 0.0207  | 0.0048  | 0.0037  | 0.0033  | 0.0032  | 0.0039  | 0.0065  | 0.0096  | 0.060   |
|   b   | 0.9587  | 0.9586  | 0.9586  | 0.9586  | 0.9585  | 0.9582  | 0.9581  | 0.9579  | 0.9581  |
|  err  | 4.28e-07| 7.55e-07| 1.08e-06| 1.29e-06| 1.49e-06| 2.63e-06| 3.78e-06| 5.33e-06| 7.30e-06|

Table 4. Values of $\alpha$, $m$, and $D$ with the corresponding error for 3D medium disorder.

|   W   | 1       | 1.5     | 2       | 2.5     | 3       | 3.5     | 4       | 4.5     | 5       |
|-------|---------|---------|---------|---------|---------|---------|---------|---------|---------|
|   α   | 1.15    | 0.78    | 0.64    | 0.56    | 0.86    | 0.59    | 0.67    | 0.60    | 0.30    |
|   m   | 0.0214  | 0.0094  | 0.0117  | 0.0147  | 0.0755  | 0.0419  | 0.0750  | 0.0966  | 0.0451  |
|   b   | 0.9580  | 0.9558  | 0.9543  | 0.9561  | 0.9506  | 0.9573  | 0.9482  | 0.9387  | 0.9232  |
|  err  | 7.49e-06| 5.38e-05| 7.38e-05| 9.57e-05| 0.0001  | 0.0005  | 0.0003  | 0.0008  | 0.0009  |

Table 5. Values of $\alpha$, $m$, and $D$ with the corresponding error for 3D strong disorder.

|   W   |   10    |   15    |   20    |    25    |    30    |   35    |    40    |
|-------|---------|---------|---------|----------|----------|---------|----------|
|   α   |  0.17   |  0.23   |  0.3    |  **0.01**|  **0.04**|  0.72   |  0.13    |
|   m   | 0.3218  | 0.9284  | 2.1973  | 13.1800  |  4.2850  | 15.2130 | 1.9124   |
|   b   | 0.7220  | 0.4783  | 0.0567  | -11.9390 | -3.1841  | 0.0623  | -0.7145  |
|  err  | **0.0073** | **0.0325** | **0.0557** | **0.0585** | **0.0751** | **0.0710** | **0.1011** |

## 6. Conclusion and application to material science

A mathematical formulation of the spectral approach to delocalization was introduced along with a physical interpretation in the context of the classical model of Edward and Thouless. We believe this new method is an important contribution to localization theory because it can be applied to infinite systems of any dimension without the use of periodic boundary conditions or scaling. Initial numerical simulations indicate that delocalization, characterized by extended states, will occur for $W \leq 0.6$ in 2D and for $W \leq 5$ for 3D systems. These results were determined form linear regression analysis of polynomial fitting to the time evolution of the distance parameter. The acceptable amount of error in our analysis is $err \leq 0.001$, where $err$ is the norm of the residuals.

Identifying precise values of the disorder at the transition points in both 2D and 3D cases is beyond the scope of this paper. However, the spectral method and the analysis presented here can be used to define the point where pure extended states cease to exist, whereas the dynamical approach can identify the point where transition to pure localization occurs. In this sense, both methods can be applied to the same system to study the transition interval, where, as mentioned above, the possible solutions in this interval are neither localized nor extended. The connection between the transition states and wave-particle duality makes the transition region an interesting subject for both mathematicians and theoretical physicists.

The spectral method can also be employed in material science where accurate results in 1D and 2D disordered systems are crucial for the study of quasi-1D graphene nanotubes and graphene sheets. Currently, both localization and transport have been shown to exist in graphene-based materials for weak disorder [33]–[36]. The transition from localized to extended states is still highly debated on theoretical grounds and there is disagreement on the value of the minimal conductivity [32]. Additionally, the localization lengths in graphene have been shown to be independent of size of the system for various strengths of the disorder [31], which is in contrast to scaling theory [18]. Detailed numerical and experimental study employing the spectral approach can greatly contribute to such problems allowing more accurate predictions for the behavior of graphene-based materials.